\documentclass[runningheads]{llncs}

\usepackage{graphicx}
\usepackage[ruled,vlined,linesnumbered]{algorithm2e}
\usepackage{amsmath} 
\usepackage{amssymb}
\usepackage{mathrsfs}

\begin{document}
\title{An almost linear time complexity algorithm for the Tool Loading Problem
}
%
%
\author{Mikhail Cherniavskii
\and
Boris Goldengorin
}
\authorrunning{M. Cherniavskii and B. Goldengorin}
%
\institute{Department of Discrete Mathematics,\\
Moscow Institute of Physics and Technology, Moscow, Russia
\email{cherniavskii.miu@phystech.edu}, \email{goldengorin.bi@mipt.ru}}
\maketitle              
\begin{abstract}
As shown by Tang, Denardo \cite{TangDenardo1} the job Sequencing and tool Switching Problem (SSP) can be decomposed into the following two problems. Firstly, the Tool Loading Problem (TLP) - for a given sequence of jobs, find an optimal sequence of magazine states that minimizes the total number of tool switches. Secondly, the Job Sequencing Problem (JeSP) – find a sequence of jobs minimizing the total number of tool switches. 
Published in 1988, the well known Keep Tool Needed Soonest (KTNS) algorithm for solving the TLP has time complexity $O(mn)$. Here $m$ is the total number of tools necessary to complete all $n$ sequenced jobs on a single machine. A tool switch is needed since the tools required to complete all jobs cannot fit in the magazine whose capacity $C < m$. We hereby propose a new Greedy Pipe Construction Algorithm (GPCA) with time complexity $O(Cn)$. Our new algorithm outperforms KTNS algorithm on large-scale datasets by at least an order of magnitude in terms of CPU times.

\end{abstract}
\section{Introduction}\label{sec:Introduction}
The advertisements which need to be delivered to the residents of the Netherlands must first be printed by advertisers. Then, depending on the postal code, several companies sort and pack the brochures in plastic bags (seal bags) which are delivered to all interested residents. Sorting and packing of brochures is carried out on conveyors equipped with magazines with a maximum capacity of $C \leq 32$ brochures. In the Netherlands, there are about ten thousand postal codes, $n=10^4$. The number of various brochures $m>10^3$ presenting the necessary goods to residents is estimated to be several thousand. Thus, an application of the Keep Tool Needed Soonest (KTNS) algorithm Tang, Denardo \cite{TangDenardo1} with time complexity $O(mn)$ requires at least $mn= 10^3 \times 10^4 = 10^7$ elementary operations. Here $m$ is the total number of tools (in our example, the total number of brochures necessary to complete all $n$ sequenced jobs (in our example, the total number of different zip codes) on a single machine. Compared to our Greedy Pipe Construction Algorithm (GPCA) with time complexity, $O(Cn) \approx C\times n= 32 \times 10^4 = 3.2\times10^5$ the GPCA saves at least an order of elementary operations. 

In this paper, we outline our GPCA and prove its correctness, including the time complexity $O(Cn)$.  Note that $C$ is the conveyor's magazine capacity just to remind its informal meaning with the purpose to exclude a standard interpretation of $C$ in algorithmic time complexity as an arbitrary constant. As far as we are aware, in most production lines (systems) $C\leq 32$ \cite{Crama1994,TangDenardo1,TangDenardo2}.

In manufacturing seal bags, processing time is important. When the production of packages moves from one postal code to another, the required set of brochures (tools) is changed. Thus, if the conveyor magazine does not have the brochures needed for the next postal code, the production will be interrupted to switch the brochures in the magazine. Our goal is to reduce the downtime by minimizing the total number of brochure switches. Such a problem can be formulated in terms of the well-known combinatorial optimization problem - the Job Sequencing and tool Switching Problem (SSP) (see [1-8]). Informally, SSP can be formulated as follows. We are given $n$ jobs (postal codes), $m$ tools (brochures), magazine capacity $C$, and a set of tools $T_i$ necessary to complete each job $i = 1,\dots , n$, i.e. $|\bigcup_{i=1}^{n} T_i| = m$. To solve SSP we need to find a job sequence i.e. to solve the Job Sequencing Problem (JeSP) and for the given sequence of jobs we need to find a tool loading strategy minimizing the total number of switches, i.e. to solve the Tool Loading Problem (TLP). 

For the state of the art and a recent overview of the models and algorithms for solving the TLP and JeSP we refer to \cite{Calmels2019}. 
Here the Tool Loading Problem (TLP) - for a given sequence of jobs, find the minimum number of tool switches, necessary to complete all $n$ jobs, represented by a sequence of magazine's states, containing the corresponding tools; the Job Sequencing Problem (JeSP) - find an optimal sequence of magazine states that minimizes the total number of tool switches. 

The purpose of our paper is to outline our Greedy Pipe Construction Algorithm (GPCA) for solving the TLP.
Informally, the TLP can be stated as follows. At the time corresponding to each job, the magazine contains tools that needed to complete the corresponding job. Thus, at each moment of time, some magazine slots will be occupied by the tools needed for the job. The goal of TLP is to fill the remaining empty slots with tools to minimize the total number of tool switches. Ghiani et al. \cite{GhianiGrieco2010} designed the Tailored KTNS procedure to find a lower bound on the total number of switches between two jobs for SSP. As far as we are aware, our GPCA is the first algorithm returning both an optimal solution and its optimal value to the TLP with $O(Cn)$ time complexity.

Let's consider an Example~\ref{example:1} of solving the TLP by the Keep Tool Needed Soonest (KTNS) algorithm proposed in Tang, Denardo \cite{TangDenardo1}. The KTNS algorithm is well known for more than $33$ years (see \cite{Ahmadi2018,Calmels2019,Catanzaro2015,CherniavskiiGoldengorin,Crama1994,GhianiGrieco2010,Mecler2021,TangDenardo1,TangDenardo2}).
\begin{example}
$ $\newline
\begin{minipage}{0.3\textwidth}
\includegraphics[width=\linewidth]{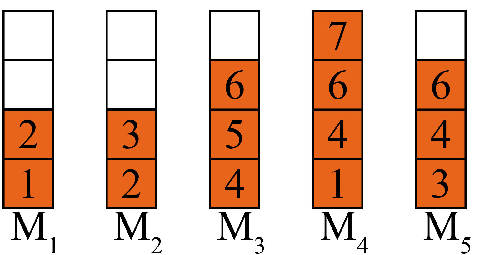}
\end{minipage}%
\hfill
\begin{minipage}{0.68\textwidth}
We are given a set of five jobs represented by the corresponding sets of tools $ T_1 = \{1,2 \} $, $ T_2 = \{2,3 \} $, $ T_3 = \{4,5,6 \} $, $ T_4 = \{1, 4,6,7 \} $, $ T_5 = \{3,4,6 \} $ needed to process each of the given jobs with magazine's capacity $C=4$ . We assume that the tools in the columns are unordered. The columns in {\it Example 1} represent magazine states $M_{1},M_{2},M_{3},M_{4},M_{5}$ and are associated with the corresponding moments $ 1, 2, 3, 4, 5$. Let $s$ be the number of tool switches. We initialize the variable $s:=0$.
\end{minipage}

\noindent \begin{minipage}{0.3\textwidth}
\includegraphics[width=\linewidth]{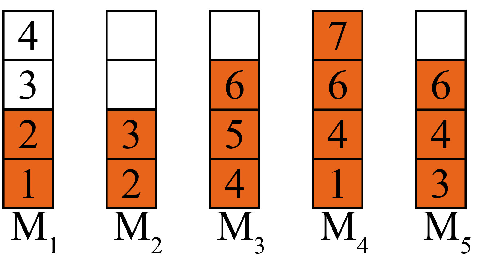}
\end{minipage}%
\hfill
\begin{minipage}{0.68\textwidth}
First, we need to fill two empty slots in the first column $M_1$ with tools that will be needed soonest. Tool 3 is needed at time moment 2, tools 4, 5, 6 are needed at moment 3, tool 7 is needed at moment 4. We do not consider tools 1 and 2, since they are already in the magazine. We fill one empty slot with tool 3, since it will be needed foremost. The second empty slot can be filled with any of the tools 4, 5, 6. Let's choose tool 4.
\end{minipage}

\noindent \begin{minipage}{0.3\textwidth}
\includegraphics[width=\linewidth]{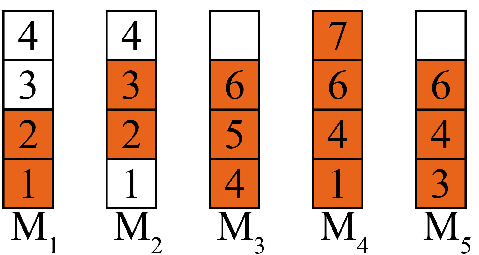}
\end{minipage}%
\hfill
\begin{minipage}{0.68\textwidth}
We continue to fill two empty slots in the column $M_2$. To do this, we need to choose two tools from the previous state $M_1$  of the magazine that needed soonest and are not yet in the magazine state $M_2$. $M_1 \backslash M_2 = \{ 1,4 \}$. If there are more than two tools then we choose among them the two that are needed first. We do not need to make a choice since there are only two tools. We fill the remaining empty slots with tools 1, 4. We can compute the number of tool switches between $M_1$ and $M_2$ since they are full. No switch is needed for transition from the magazine state $M_1$ to state $M_2$ since $M_1$ and $M_2$ are the same. Now $s:= s + |M_2 \backslash M_1| = 0 + |\{1,2,3,4\} \backslash \{ 1,2,3,4 \}|=0$.
\end{minipage}

\noindent \begin{minipage}{0.3\textwidth}
\includegraphics[width=\linewidth]{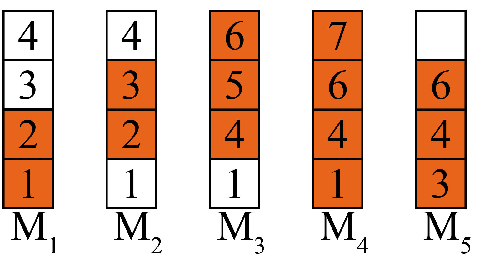}
\end{minipage}%
\hfill
\begin{minipage}{0.68\textwidth}
Next we fill the remaining empty slot in the column $M_3$. To do this, we need to choose one tool from the previous state of the magazine $M_2$ that is needed soonest and is not yet in the magazine state $M_3$. $M_2 \backslash M_3 = \{ 1,2,3 \}$. Tool 1 is needed at time 4, tool 2 will not be needed again, and tool 3 is needed at time 5. We now fill the empty slot with the tool that is needed soonest, which is $1$. We can now compute the number of tool switches between $M_2$ and $M_3$ because they are full. $s:= s + |M_3 \backslash M_2| = 0 + |\{1,4,5,6\} \backslash \{ 1,2,3,4 \}|=0 + |\{ 5,6 \}| = 2$. Thus, for transition from magazine state $M_2$ to state $M_3$ tools $2,3$ are switched by $5,6$.

\end{minipage}

\noindent \begin{minipage}{0.3\textwidth}
\includegraphics[width=\linewidth]{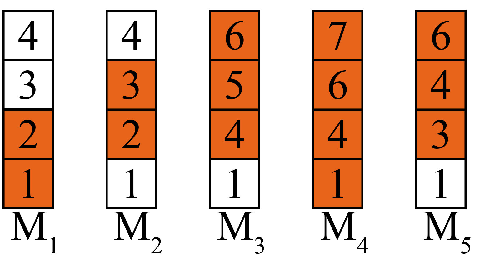}
\end{minipage}%
\hfill
\begin{minipage}{0.68\textwidth}
Since the magazine state $M_4$ has no empty slots, there is nothing to fill. $s:=s+|M_4 \backslash M_3| = 2 + |\{1,4,6,7\} \backslash \{ 1,4,5,6\}| = 2 + |\{7\}| = 3$. For the next transition from magazine state $M_3$ to state $M_4$ tool $5$ is replaced by tool $7$. Let us move on to filling the empty slot in $M_5$. To do this, we need to choose one tool that is needed soonest from the previous state $M_4$ and is not yet in $M_5$. $M_4 \backslash M_5 = \{ 1,6,7 \}$. Tools 1, 6, and 7 will not be needed again so we chose an arbitrary tool, i.e. tool 1. $s:=s+|M_5 \backslash M_4| = 3 + |\{1,3,4,6\} \backslash \{1,4,6,7\}| = 3 + |\{3\}| = 4$. All magazine slots are now filled. In conclusion, the total number of tool switches is $s = |M_2 \backslash M_1| + |M_3 \backslash M_2|+|M_4 \backslash M_3| +|M_5 \backslash M_4| = 0 + 2 + 1 + 1= 4$ terminating KTNS algorithm. Further in Example~\ref{example:2} we solve the same problem by our algorithm GPCA to show differences between KTNS and GPCA.
\end{minipage}\label{example:1}
\end{example}

The article is organized as follows. In Section 2 we introduce formal notation formulating SSP and TLP. In Section 3 we propose the GPCA algorithm for computing the objective function value of SSP and ToFullMag algorithm which is processed after GPCA and returns an optimal solution to the TLP. Further, we formulate theorems to justify the correctness of GPCA and its applicability for computing the SSP objective function value. Next, we illustrate the steps of our GPCA by means of a numerical example including GPCA's efficient implementation and its time complexity. Section 4 presents our computational results. In Section 5 we discuss some conclusions and future research directions.
\section{Problem formulation}
	Let's introduce the basic notation for the job Sequencing and tool Switching Problem (SSP) and further illustrate with examples.
	$T = \{ 1,2, \dots ,m \}$ is the set of tools, where $ m $ is the number of tools.
	$J=\{1,2 \dots n \}$ is the set of jobs, where $ n $ is the number of jobs.
	$T_i \subset T$ is the set of tools needed to complete the job $ i \in J $. $ C <m $ is the capacity (number of slots) of the magazine.
	$M_i \subset T$ is the state of the magazine when the job $ i $ is processed, i.e. the set of tools located in the magazine to process the job $i \in J$, $|M_i|=C$, $T_i \subseteq M_i$.
	$\boldsymbol{M} = \{ M = (M_{1}, \dots , M_{n}): T_i \subseteq M_i, |M_i|=C, i \in J \}$  is the set of all sequences of magazine states such that jobs can be processed in order $(1,2, \dots , n)$. 
	$switches(M) = |M_2 \backslash M_1| + |M_3 \backslash M_2| + \dots |M_n \backslash M_{n-1}|  = \sum_{i=1}^{n-1} (C - |M_{{i}} \cap M_{{i+1}}|)$ is the number of switches for the sequence of magazine states, where $M \in \boldsymbol{M}$. The Tool Loading Problem (TLP) is the problem of finding a sequence of magazine states $M^{*}$ that minimizes the number of switches for the given set of sequenced tools $T_1,\dots , T_n$ and magazine capacity $C$, i.e. $M^{*} \in argmin\{ switches(M): M \in \boldsymbol{M}\}$.

\begin{figure}[ht]
\begin{minipage}[t]{.45\linewidth}
  \includegraphics[width=\linewidth]{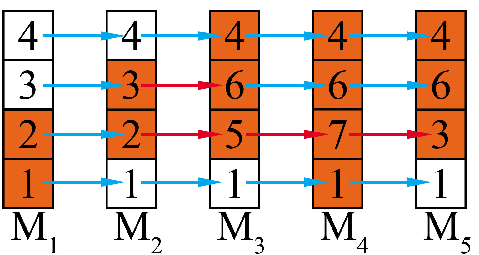}
  \caption{TLP solution with sorting\\ ready for production.}
  \label{fig:1}
\end{minipage}
\hfill
\begin{minipage}[t]{.45\linewidth}
  \includegraphics[width=\linewidth]{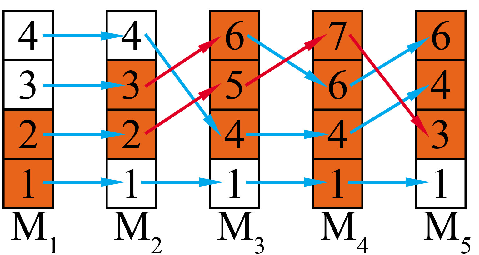}
  \caption{TLP solution without sorting.}
  \label{fig-2}
\end{minipage}
\end{figure}
Fig.~\ref{fig:1} shows a solution for Example~\ref{example:1} where red arcs indicate switches while blue arcs indicate that the tool is scheduled both in the current state of the magazine and in the next one, i.e. no switch is required. In Fig.~\ref{fig:1} the columns corresponding to the magazine states are sorted so that the picture contains information about which slot will contain which tool at any given time. For example, the bottommost slot will have states $(1,1,1,1,1)$, the second slot $(2,2,5,7,3)$, the third slot $(3,3,6,6,6)$, and the fourth slot $(4,4,4,4,4)$. Such a picture fully reflects which tool should be loaded to which slot in the production process. In the $TLP$ we deal with an unordered set of tools at each moment within the magazine, minimizing the total number of switches. In other words the contents of columns in Fig.~\ref{fig-2} are sorted arbitrarily, but to use a specific order we draw the columns sorted in ascending order of tool numbers and consider $M_1,M_2,\dots,M_n$ unordered sets. If at a certain moment the tool is needed to complete a scheduled job, then it is shown with an orange background. If the tool is not needed for the sequenced job at a given moment, then it is shown with a white background, i.e. $T_i$ marked in orange, 
$M_i \backslash T_i$ marked in white. Thus $T_1 = \{1,2\}$, $M_1 \backslash T_1 = \{3,4\}$, $T_2 = \{2,3\}$, $M_2 \backslash T_2 = \{1,4\}$, $T_3 = \{4,5,6\}$, $M_3 \backslash T_3 = \{1\}$, $T_4 = \{1,4,6,7\}$, 
$M_4 \backslash T_4 = \varnothing$, 
$T_5 = \{3,4,6\}$, $M_5 \backslash T_5 = \{1\}$.
\begin{figure}[h]
\begin{center}
  \includegraphics[width=.4\linewidth]{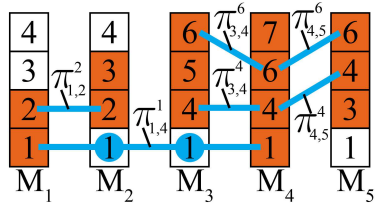}
  \caption{Examples of pipes.}
  \label{fig-3}
\end{center}
\end{figure}
\newline Let us introduce a {\it pipe} $\pi_{s, e}^{ t}$ tracing the tool $t$ from starting moment $s$ through ending moment $e$. $\mathscr{P}(M) = \{\pi_{s,e}^{t}: t \in (T_{s} \cap T_{e}) \backslash (\bigcup_{i=s+1}^{e-1} T_{i}), t \in  \bigcap_{i=s+1}^{e-1} M_{i},s < e \}$ denotes the set of all pipes $\pi_{s, e}^{ t}$ such that the tool $t$ is used for jobs at moments $ s, e $, and not used for jobs at intermediate moments $s+1, \dots , e -1$. Anyway, the tool $t$ is present in the magazine at all intermediate moments $s+1, \dots , e-1$, despite that this tool $t$ is not used for any of currently processed jobs.
The main object of this article will be a {\it pipe}. Fig.~\ref{fig-3} shows examples of pipes. Informally, a pipe is the saving of the tool $ t $ from the moment $ s $, where it was used for a job (marked in orange) until the moment $ e $, where it will again be used for another job (marked in orange), but at intermediate time moments $ s + 1, \dots , e-1 $ the tool $t$ is not needed for any sequenced job (marked in white). For example, $ \pi_{1, 4}^{ 1} $ saves tool $ 1 $ from time $ 1 $ through time $ 4 $ even though it is not used for any job at times $ 2, 3 $. Note that $ \pi_{1, 2}^{ 2}$ is also a pipe, without intermediate times between the times $ 1, 2 $. Remaining pipes are $ \pi_{3, 4}^{ 4}, \pi_{3, 4}^{ 6}, \pi_{4, 5}^{ 4}, \pi_{4, 5}^{ 6}$.
\section{GPCA policy}

Tang, Denardo \cite{TangDenardo1} suggested solving TLP by Keep  Tool  Needed  Soonest (KTNS) algorithm with $ O (mn) $ time complexity. 
Note that $m = | \bigcup_{i =1}^{n} T_{i} |$ and $|T_i|\leq C$, then in the worst case $m= C n $, thus the time complexity of KTNS is $O(C n^2)$. In this paper, we propose a new Greedy Pipe Construction Algorithm (GPCA) which computes the TLP objective function value with time complexity $ O (Cn) $. 
We also propose an algorithm ToFullMag such that consistent termination of GPCA and ToFullMag returns an optimal solution to the TLP with time complexity $O(Cn)$. An illustration of all GPCA and ToFullMag steps in Example~\ref{example:2} is provided.

\begin{algorithm}[ht]\label{algorithm:1}
\SetAlgoLined
\DontPrintSemicolon
\SetKw{KwGoTo}{go to}
	$pipes\_count:= 0$\;	
	$M_1: = T_1, M_2: = T_2, \dots , M_n: = T_n$\\
	\For{$e=2,\dots,n$}{
		$candidates:= \{\pi_{s,e}^{t} : \exists \ s<e: t \in (T_{{s}} \cap T_{{e}}) \backslash (\bigcup_{i=s+1}^{e-1} T_{i}) \}$  $/^{*}$\textit{the set of all possible pipes with endings at time $ e $.} $^{*}/$\;  	
		\For{$\pi_{s,e}^{t} \in candidates$ $/^{*}$\textit{enumerate in any order} $^{*}/$}{
			\If{$\forall i \in \{s+1, \dots ,e-1\} \ |M_{{i}}| < C$ $/^{*}$\textit{i.e. there are enough empty slots to build $\pi_{s,e}^{t}$} $^{*}/$}{

				add $ t $ to $M_{{s+1}}, M_{{s+2}}, \dots , M_{{e-1}} $ $/^{*}$\textit{i.e. build a pipe $\pi_{s,e}^{t}$} $^{*}/$\;
				$pipes\_count:=pipes\_count+1$\;
			}
		}
    }
  \caption{GPCA }
\end{algorithm}

Recall that $\mathscr{P}(M)$ is the set of all pipes in the sequence of magazine states $M$. $\boldsymbol{M}$ is the set of all possible sequences of the magazine states such that given jobs can be completed in sequence $1,\dots , n$. $switches(M)$ denotes the number of tool switches in $M$. Theorem~\ref{theorem-1} states that the TLP objective function equals to $\sum_{i=1}^{n}|T_i| -C - max\{|\mathscr{P}(M)|: M \in \boldsymbol{M}|\}$, i.e. the minimum number of switches that is required to complete a sequence of jobs equals to $\sum_{i=1}^{n}|T_i| -C - max\{|\mathscr{P}(M)|: M \in \boldsymbol{M}|\}$.  Note that magazine's capacity $C$ and tool sets $T_1 , \dots , T_n$ are given and fixed, which implies that minimizing the number of switches $switches(M)$ is equivalent to maximizing the number of pipes $|\mathscr{P}(M)|$. The main point of Theorem~\ref{theorem-1} is that in order to find the minimum number of switches, it suffices to know the maximum number of pipes.

\begin{theorem}\label{theorem-1}
Let $ C $ be the capacity of the magazine, $ T_1, \dots T_n $ are the required sets of tools for jobs $ 1, \dots, n $, then

$
min\{switches(M):M \in \boldsymbol{M}\} = \sum_{i=1}^{n}|T_i| -C - max\{|\mathscr{P}(M)| :M \in \boldsymbol{M}\}.$
\end{theorem}

Statement 1 of Theorem~\ref{theorem-2} claims that GPCA constructs the maximum possible number of pipes, i.e. $GPCA(T_1,\dots ,T_n;C)=max\{|\mathscr{P}(M)|: M \in \boldsymbol{M}|\}$ and therefore, according to Theorem 1, the TLP objective function can be computed as $\sum_{i=1}^{n}|T_i|  -C  - GPCA(T_1,\dots ,T_n;C)$.
Statement 2 claims that sequential processing of the GPCA and ToFullMag algorithms returns an optimal solution of TLP.
\begin{theorem}\label{theorem-2}
Let $ C $ be the capacity of the magazine, $ T_1, \dots T_n $ are the required sets of tools for jobs $ 1, \dots, n $, then
\begin{enumerate}
\item $min\{switches(M): M \in \boldsymbol{M}|\}  = \sum_{i=1}^{n}|T_i|  -C  - GPCA(T_1,\dots ,T_n;C)$
\item $ToFullMag(GPCA(T_1 ,\dots T_n;C)) \in argmin\{switches(M): M \in \boldsymbol{M}\}$
\end{enumerate}
\end{theorem}
\begin{example}
$ $\newline
\begin{minipage}{0.3\textwidth}
\includegraphics[width=\linewidth]{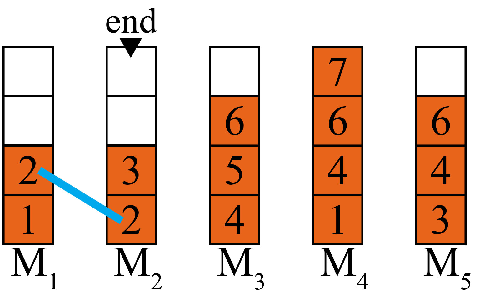}
\end{minipage}%
\hfill
\begin{minipage}{0.68\textwidth}
Let us solve the same problem as in Example~\ref{example:1} using GPCA and ToFullMag algorithms.
    The first value of the variable $ e = 2 $. Let's find all pipes that end at time 2. There are two tools in the magazine state $M_2$ they are tool $2$ and tool $3$. We have to find the last point in time when each of the tools was used. Tool $2$ used at time $1$ thus we have found a pipe $\pi_{1,2}^{2}$. The tool $3$ has never been used before the moment $2$, hence a pipe $\pi_{s,2}^{3}$ does not exist for any $s<e=2$. There is only one candidate for construction, that is $\pi_{1,2}^{2}$. No empty slots are required to build pipe $\pi_{1,2}^{2}$, therefore $\pi_{1,2}^{2}$ will be constructed (marked in blue in the figure).
\end{minipage}
\begin{minipage}{0.3\textwidth}
\includegraphics[width=\linewidth]{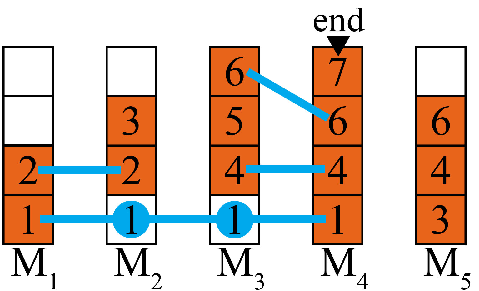}
\end{minipage}%
\hfill
\begin{minipage}{0.68\textwidth}
Let $ e = 3 $. $M_{3}=\{4,5,6\}$. 
Tools $4,5,6$ have never been used before the moment $3$, hence no pipe $\pi_{s,3}^{t}$ exists for any $s<e=3$, $t \in \{4,5,6\}$. 
Let $ e = 4 $. $M_{4}=\{1,4,6,7\}$. 
Tool $1$ used at time $1$ thus we have found a pipe $\pi_{1,4}^{1}$.
Tools $4,6$ used at time $3$ then we have found pipes $\pi_{3,4}^{4}$ and $\pi_{3,4}^{6}$. 
The tool $7$ has never been used before the moment $4$, then a pipe $\pi_{s,2}^{7}$ does not exist for any $s<e=4$. 
Candidates for construction are: $\pi_{1,4}^{1}$, $\pi_{3,4}^{4}$, $\pi_{3,4}^{6}$. 
Construction of pipe $\pi_{1,4}^{1}$ requires one empty slot at moments 2,3. 
There are two empty slots at time $2$ and one empty slot at time $3$, then $\pi_{1,4}^{1}$ will be constructed. 
No empty slots are required to build pipes $\pi_{3,4}^{4}$, $\pi_{3,4}^{6}$, then $\pi_{3,4}^{4}$, $\pi_{3,4}^{6}$ will be constructed.
\end{minipage}
\begin{minipage}{0.3\textwidth}
\includegraphics[width=\linewidth]{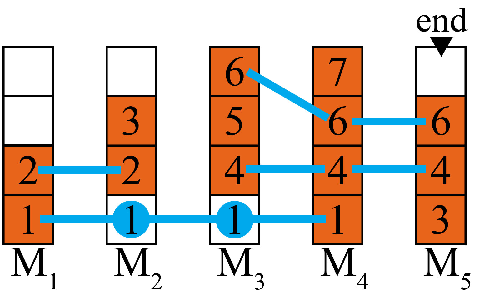}
\end{minipage}%
\hfill
\begin{minipage}{0.68\textwidth}
Let $ e = 5 $. $M_{5}=\{3,4,6\}$. 
Tool $3$ used at time $2$ then we have found a pipe $\pi_{2,5}^{3}$.
Tools $4,6$ used at time $4$ then we have found pipes $\pi_{4,5}^{4}$ and $\pi_{4,5}^{6}$.
Candidates for construction are: $\pi_{2,5}^{3}$, $\pi_{4,5}^{4}$, $\pi_{4,5}^{6}$. 
Construction of pipe $\pi_{2,5}^{3}$ requires one empty slot at moments 3,4. There are no empty slots at time 3, then $\pi_{2,5}^{3}$ will not be constructed. 
No empty slots are required to build pipes $\pi_{4,5}^{4}$, $\pi_{4,5}^{6}$, then $\pi_{4,5}^{4}$, $\pi_{4,5}^{6}$ will be constructed.
Now the GPCA is terminated. $ 6 $ pipes were built, then according to Theorem~\ref{theorem-1} we have found the total number of tool switches
$min\{ switches(M): M \in \boldsymbol{M} \} = \sum_{i=1}^{n}|T_i| -C -6 = (2+2+3+4+3) - 4 - 6 = 4$.
\end{minipage}
\begin{minipage}{0.3\textwidth}
\includegraphics[width=\linewidth]{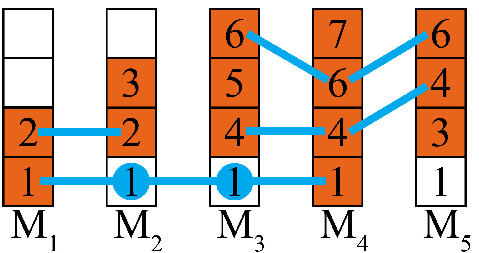}
\end{minipage}%
\hfill
\begin{minipage}{0.68\textwidth}
Note that there are four empty slots after GPCA execution. 
   Let's fill those empty slots by ToFullMag algorithm, which fills empty slots without increasing of number of switches, so the total number of switches will be $4$.
   At the first stage, ToFullMag enumerates through pairs $(M_{1},M_{2}), (M_{2},M_{3}), \dots , (M_{4}, M_{5})$. If in the pair $(M_{i},M_{i+1})$ the second element $M_{i+1}$ has empty slots, i.e. $|M_{i+1}|<C$ and there is a tool that exists in $M_{i}$ and doesn't exist in $M_{i+1}$, then such a tool is added to $M_{i+1}$. In our example, at the first stage, only one empty slot will be filled. Let's consider the pair $(M_{4},M_{5})$. $M_5$ has one empty slot, $M_{4} \backslash M_{5} = \{1,4,6,7\} \backslash \{3,4,6 \}= \{1,7\}$, then we can fill empty slot by tool $1$ or tool $7$. Let us chose tool $1$. So the final state of the magazine at the time $5$ is $M_{5}=\{1,3,4,6\}$. Note that adding tool 1 to $M_5$ did not increase the number of switches because tool 1 presents in $M_4$.
\end{minipage}

\begin{algorithm}[ht]\label{algorithm:2}
\SetAlgoLined
\DontPrintSemicolon
\SetKw{KwGoTo}{go to}
$M$ is computed by $GPCA(T_1, \dots , T_n; C)$\;
$used[t] := 0$ for all $t \in M$\;
	\For{$(i,j)=(1,2),(2,3),\dots,(n-1,n), (n,n-1),(n-1,n-2),\dots (2,1)$}{
		$used[t]:=1$ for all $t \in M_{j}$\;
		\For{$t \in M_{i}$}{
			\If{$used[t]==0$ $and$ $|M_{j}|<C$}{
				$M_{j}:= M_{j} \cup \{t\}$\;
			}
    	}
    	$used[t]:=0$ for all $t \in M_{j}$\;
    }
  \caption{\textit{ToFullMag} }
  
\end{algorithm}

\noindent
\begin{minipage}{0.3\textwidth}
\includegraphics[width=\linewidth]{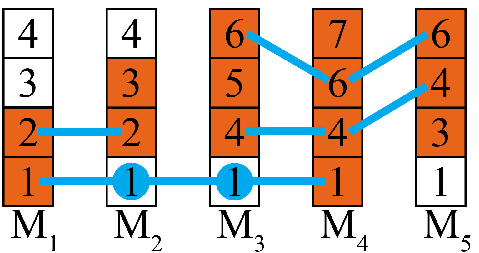}
\end{minipage}%
\hfill
\begin{minipage}{0.68\textwidth}
At the second stage, ToFullMag enumerates through pairs $(M_{5},M_{4})$, $(M_{4},M_{3}), \dots , (M_{2}, M_{1})$. 
And same as in the first stage, if in the pair $(M_{i},M_{i+1})$ the second element $M_{i+1}$ has empty slots i.e. $|M_{i+1}|<C$ and there is a tool that exists in $M_{i}$ and doesn't exist in $M_{i+1}$, then such a tool is added to $M_{i+1}$. We skip pairs $(M_{5},M_{4}),(M_{4},M_{3})$ because $|M_{4}|=C$ and $|M_{3}|=C$. Let's consider the pair $(M_{3},M_{2})$. $M_2$ has one empty slot, $M_{3} \backslash M_{2} = \{1,4,5,6\} \backslash \{1,2,3 \}= \{4,5,6\}$, then we can fill empty slot by tool $4$ or tool $5$ or tool $6$. Let us chose tool $4$. So the final state of the magazine at the time $2$ is $M_{2}=\{1,2,3,4\}$. Let's consider the pair $(M_{2},M_{1})$. $M_{1}$ has two empty slots, $M_{2} \backslash M_{1} = \{1,2,3,4\} \backslash \{1,2 \}= \{3,4\}$, then we can fill empty slots by tools $3$ and $4$. So the final state of the magazine at the time $1$ is $M_{1}=\{1,2,3,4\}$.
Note that adding tool 4 to $M_1$ and $M_2$ didn't increase the number of switches, but only changed the instant of loading  of tool $4$ from instant 3 to instant 1. Adding tool $3$ to $M_1$ did not increase the number of switches, but only changed the moment of loading of tool $3$ from time 2 to time 1. Note that there are exactly 4 switches: $|M_{2} \backslash M_{1}| + |M_{3} \backslash M_{2}| + |M_{4} \backslash M_{3}| + |M_{5} \backslash M_{4}| = 0 + 2 + 1 + 1=4$. So we got the same solution $M_1,\dots ,M_n$ with Example~\ref{example:1} and the same number of switches.
\end{minipage}
\label{example:2}
\end{example}

\begin{algorithm}[ht]\label{algorithm:3}
\SetAlgoLined
\DontPrintSemicolon
\SetKw{KwGoTo}{go to}
	$pipes\_count:= 0$, $last\_full := 0$\;
	$last\_seen[t]:= 1$ for all $t \in T_{{1}}$, $-1$ for all $t \in T \backslash T_{1}$\;
	$M_i:=T_i$ for all $i=1, \dots , n$\;

	\For{$e=1,\dots,n$}{
		\For{$t \in T_{{e}}$}{
			\If{$last\_full \leq last\_seen[t]$}{
				$pipes\_count := pipes\_count + 1$\;
				\For{$i = last\_seen[t]+1, last\_seen[t]+2, \dots , e-1$}{
					$M_i:= M_i \cup \{ t \}$\;
					\If{$|M_i|==C$}{
						$last\_full := i$\;
					}
				}
			}
			$last\_seen[t] := e$\;
    	}
		\If{$|M_{e}|==C$}{
			$last\_full := e$\;	
		}
    }
  \caption{\textit{GPCA implementation} }
\end{algorithm}

The algorithm ToFullMag takes as an input a sequence of magazine states $M_1, \dots , M_n$ obtained by using GPCA and fills the remaining empty slots without increasing the number of switches.

Thus, if $M^{*}=ToFullMag(GPCA(T_1,\dots ,T_n;C))$, then $M$ is an optimal TLP solution, i.e. $M^{*} \in  argmin\{ switches(M): M \in \boldsymbol{M}\}$.

Let's consider the time complexity of ToFullMag.
The loop in line 1 does $m$ iterations, but $m = |\bigcup_{i=1}^{n} T_i| \leq Cn$. The loop in line 2 does at most $2 n$ iterations, the loops in lines 3, 4, and 9 take at most $C$ iterations, hence the complexity of ToFullMag is $O(Cn)$.

The Algorithm~\ref{algorithm:3} uses the Algorithm~\ref{algorithm:1} property, which allows us to iterate over tools in any order (see Algorithm 1 line 5). Let's create an array $ last\_seen $, where $ last\_seen[t]$ is the last moment in time when tool $t$ was needed for a job. The variable $ last\_full $ is equal to the last moment in time when the magazine was full, so if $last\_full> last\_seen [t]$, then the pipe $ \pi_{last\_seen[t],e}^{t} $ cannot be built, since there are not enough empty slots. But if $ last\_full \leq last\_seen [t] $, then the algorithm builds the pipe $\pi_{last\_seen[t],e}^{t} $.

Let us analyze the time complexity of the Algorithm~\ref{algorithm:3}. The time complexity of line $1$ is $O(1)$, line $2,3$ is $m \leq Cn$. The loop in line $ 4 $ does $ n-1 $ iterations, the loop in line $ 6 $ does no more than $ C $ iterations. 
Each execution of the line $ 9 $ means filling in one slot of the magazine, of which there are only $ C n $, so the line $ 9 $ will be called no more than $ C n $ times. All remaining lines have $O(1)$ complexity. From all of the above, it follows that the time complexity of GPCA is $O(Cn)$.

According to Theorem~\ref{theorem-2} and the fact that time complexities of GPCA and ToFullMag are $O(Cn)$ we can state the following theorem.
\begin{theorem}
The time complexity of TLP is $O(Cn)$.
\end{theorem}

\begin{table}[p]
\centering
\caption{Results of computational experiments for KTNS, GPCA and ToFullMag(GPCA) for $A, B, C, D$ datasets from Catanzaro et al. \cite{Catanzaro2015}, and  $F1, F2, F3$ from Mecler et al. \cite{Mecler2021}.}
\begin{tabular}{c|c|c|c|c|c|c}
\hline
dataset &   n &   m &   C &    KTNS, s &   GPCA, s &  ToFullMag(GPCA), s \\
\hline
          A1 &  10 &   10 &   4 &    1.377 &   0.268 &      0.454 \\
          A2 &  10 &   10 &   5 &    1.334 &   0.262 &      0.580 \\
          A3 &  10 &   10 &   6 &    1.215 &   0.295 &      0.671 \\
          A4 &  10 &   10 &   7 &    1.049 &   0.313 &      0.749 \\
\hline
          B1 &  15 &   20 &   6 &    5.493 &   0.513 &      0.937 \\
          B2 &  15 &   20 &   8 &    5.187 &   0.688 &      1.235 \\
          B3 &  15 &   20 &  10 &    4.554 &   0.719 &      1.453 \\
          B4 &  15 &   20 &  12 &    3.969 &   0.702 &      1.705 \\
\hline
          C1 &  30 &   40 &  15 &   34.291 &   2.531 &      4.064 \\
          C2 &  30 &   40 &  17 &   31.021 &   2.797 &      4.436 \\
          C3 &  30 &   40 &  20 &   26.690 &   2.999 &      4.873 \\
          C4 &  30 &   40 &  25 &   19.972 &   3.189 &      5.469 \\
\hline
          D1 &  40 &   60 &  20 &   82.308 &   4.327 &      6.719 \\
          D2 &  40 &   60 &  22 &   76.167 &   4.815 &      7.360 \\
          D3 &  40 &   60 &  25 &   69.728 &   5.171 &      7.889 \\
          D4 &  40 &   60 &  30 &   60.206 &   5.563 &      8.593 \\
\hline
        F1.1 &  50 &   75 &  25 &  138.200 &   6.922 &     10.156 \\
        F1.2 &  50 &   75 &  30 &  120.390 &   7.922 &     11.547 \\
        F1.3 &  50 &   75 &  35 &  105.540 &   8.267 &     12.405 \\
        F1.4 &  50 &   75 &  40 &   91.701 &   8.438 &     13.170 \\
\hline
        F2.1 &  60 &   90 &  35 &  230.450 &  11.405 &     16.703 \\
        F2.2 &  60 &   90 &  40 &  202.430 &  12.736 &     18.735 \\
        F2.3 &  60 &   90 &  45 &  176.880 &  13.328 &     19.390 \\
        F2.4 &  60 &   90 &  50 &  155.350 &  13.734 &     20.578 \\
\hline
        F3.1 &  70 &  105 &  40 &  331.310 &  15.312 &     22.390 \\
        F3.2 &  70 &  105 &  45 &  296.950 &  16.782 &     24.125 \\
        F3.3 &  70 &  105 &  50 &  270.680 &  17.655 &     25.249 \\
        F3.4 &  70 &  105 &  55 &  256.880 &  18.501 &     25.922 \\
\hline
\end{tabular}\label{table:1}
\end{table}

\begin{figure}[p]
\begin{minipage}[t]{0.95\linewidth}
  \includegraphics[width=\linewidth]{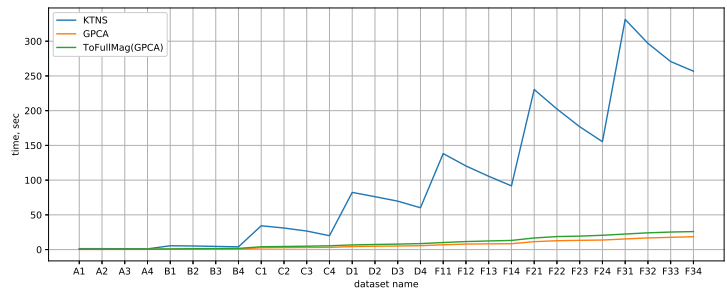}
  \caption{Comparison of KTNS, GPCA and ToFullMag(GPCA) for datasets Catanzaro et al. \cite{Catanzaro2015}, Mecler et al. \cite{Mecler2021}.}
  \label{fig-4}
\end{minipage}
\end{figure}

\section{Computational experiments}

All computations were implemented on an Intel\textsuperscript{®} Core\textsuperscript{TM} i$5$ CPU $2.60$ GHz computer with 4 GB of RAM.
GPCA and ToTfullMag were implemented in $C++$.  Mecler et al.\cite{Mecler2021} published on the GitHub repository an algorithm $HGS$ for solving SSP, in which the $KTNS$ algorithm was used to compute the objective function value. The source code of the KTNS C++ program was taken from \url{https://github.com/jordanamecler/HGS-SSP}. Catanzaro et al. \cite{Catanzaro2015} and Mecler et al. \cite{Mecler2021} datasets are also available at this link. GPCA, KTNS, ToTfullMag were compiled with g++ version 10.3.0 using $-O3$ flag. To compare the algorithms, $ 10^5 $ job sequences were generated for each dataset from Catanzaro et al. \cite{Catanzaro2015}, each dataset contains $10$ problem instances. Additionally, $2 \cdot 10^5 $ job sequences were generated for each dataset of Mecler et al. \cite{Mecler2021}, each dataset contains $5$ problem instances. Thus,  Table~\ref{table:1}, Fig.~\ref{fig-4} shows the computational time of algorithms for $ 10^6 $ sequences for each dataset. $A, B, C, D$ datasets are  generated from Catanzaro et al. \cite{Catanzaro2015} and $F1, F2, F3$ datasets borrowed from Mecler et al. \cite{Catanzaro2015}. Computational experiments were carried out according to recommendations given by Johnson \cite{Johnson2002}.

Fig.\ref{fig-4} shows that the computational time of KTNS is not monotone and decreases with an increase in the capacity of the magazine $C$. Note that for GPCA and ToFullMag(GPCA), in contrast to KTNS, the computational time increases with increasing magazine capacity.

\section{Conclusions} 
Our GPCA, ToFullMag(GPCA) algorithms outperform the KTNS algorithm by at least an order of magnitude in terms of CPU times for large-scale datasets of type F3 in Mecler et al.\cite{Mecler2021}. Also, the time complexity of our algorithms is $O(Cn)$ while KTNS has $O(mn)$. Our future studies on the replacement of KTNS (which is currently used to calculate the objective function JeSP in the overwhelming majority of articles) by GPCA will be implemented and evaluated. Note that while our new algorithms are dependent on the number of jobs and magazine capacity, they are completely independent of the total number of tools.

%
%
%

\bibliographystyle{splncs04}
\bibliography{bibliography}

\newpage

\appendix
\section{Appendix}\label{sec:PROOFS}

\subsection{Proof of Theorem 1}
Let $ \boldsymbol {L} = \{L = (L_ {1}, \dots, L_ {n}): \forall i \in J $ $ T_i \subseteq L_i, | L_i | \leq C \} $ be a magazine state sequence in which jobs can be implemented in the order $1,2, \dots n$, where empty slots are allowed by the condition $ | L_i | \leq C $. 
Note that $ \boldsymbol {M} \subseteq \boldsymbol {L} $ and sets  $\boldsymbol {L}$, $\boldsymbol {M}$, differ only in terms $| L_i | \leq C$ for $\boldsymbol{L}$ and $| M_i | = C$ for $\boldsymbol{M}$.  \\
Let $G_L=(V(G_L),A(G_L))$ denote graph where $V(G_L) = \{v_{i}^{t}: t\in L_{i}\}$, $A(G_L) = \{(v_{i}^{t},v_{i+1}^{t}): v_{i}^{t},v_{i+1}^{t} \in V\}$. $V(G_L)$ shows the content of each slot of the magazine at each moment of time, since $v_{i}^{t} \in V(G_L)$ iff a tool $t$ is contained in the magazine at the instant $i$. Note that if $(v_{i}^{t},v_{i+1}^{t}) \in A(G_L)$, then tool $t$ is not switched by another tool when transitioning from state $L_i$ to state $L_{i+1}$. 

$A(G_L)$ shows all the places where switch is not needed. If $M \in \boldsymbol{M}$, then $(v_{i}^{t},v_{i+1}^{t}) \in A(G_M)$ iff no switch of the tool $t$ at the instant $i+1$ is needed. Therefore, the number of tool switches in $M \in \boldsymbol{M}$ equals to $C (n-1) - |A|$, i.e. the number of all possible places where switch might be needed minus the number of places where switch is not needed.

For a subgraph $G$ of the graph $G_L$, let $U(G)=\{v_{i}^{t} \in V(G): t \notin T_i \}$ be the set of useless vertices, i.e. vertices $v_{i}^{t}$ such that the tool $t$ is in magazine at the time $i$, but $t$ is not needed for the job $T_i$. An example of graph $G_L$ shown in Fig. \ref{fig-2}, where set of blue arcs is $A(G_m)$, red arcs should be ignored, orange slots are $U(G_L)$, white slots are $V(G_L) \backslash U(G_L)$.
$\mathscr{H}^{post}_1(L):=\{
p_{s,e}^{t} = (v_{s}^{t},v_{s+1}^{t}, \dots v_{e}^{t}) :
t \in T_s, t \notin \bigcup_{i=s+1}^{e} T_{i}, t \in \bigcap_{i=s+1}^{e} L_{i}, \\ (t \notin L_{e+1} \ or \ e=n)
\}$ is set of paths in $G_L$, which correspond to the situation when $t$ is needed for job $T_s$, not needed for jobs $T_{s+1}, \dots T_{e}$, but was kept in states $L_{s+1}, \dots L_{e}$ and was removed at time $e+1$ or $e=n$.
$\mathscr{H}^{pre}_1(L):=\{
p_{s,e}^{t} = (v_{s}^{t},v_{s+1}^{t}, \dots v_{e}^{t}) :
t \in T_e, t \notin \bigcup_{i=s}^{e-1} T_{i}, t \in \bigcap_{i=s}^{e-1} L_{i}, (t \notin L_{s-1} \ or \ s=1)
\}$ is set of paths in $G_L$, which correspond to the situation when $t$ was not in the magazine at time $s-1$ or $s=1$ and $t$ is needed for job $T_e$, not needed for jobs $T_{s}, \dots T_{e-1}$, but was inserted into the magazine in advance at time $s$ and stayed in the magazine until the moment $e$. 
$\mathscr{H}_0(L):=\{
p_{s,e}^{t} = (v_{s}^{t},v_{s+1}^{t}, \dots , v_{e}^{t}) :
t \notin \bigcup_{i=s}^{e} T_{i}, t \in \bigcap_{i=s}^{e} L_{i}, (t \notin L_{s-1} \ or \ s=1), (t \notin L_{e+1} \ or \ e=n)
\}$ is set of paths in $G_L$, which correspond to the situation when $t$ was inserted into the magazine at the moment $s$, $t$ was in the magazine at instants $s,s+1, \dots , e$ and was removed from the magazine at the time $e+1$. However, at no point in time $s,s+1, \dots ,e$ the tool $t$ was needed for jobs, i.e. the tool $t$ was inserted into the magazine in vain. 
$\mathscr{H}_1(L):= \mathscr{H}^{pre}_1(L) \cup \mathscr{H}^{post}_1(L)$.\\
$\mathscr{P}(L):=\{
\pi_{s,e}^{t} = (v_{s}^{t},v_{s+1}^{t}, \dots , v_{e}^{t}) :
t \in T_s \cap T_e, t \notin \bigcup_{i=s+1}^{e-1} T_{i}, t \in \bigcap_{i=s+1}^{e-1} L_{i}, s < e
\}$ is the set of paths in $G_L$, which correspond to the situation when $t$ needed for jobs $T_s$, $T_e$, not needed for jobs $T_{s+1}, \dots T_{e-1}$, i.e., at the instant $s$ the tool $t$ was used to implement the job $T_s$, it was kept in the magazine in the states $L_{s+1}, \dots L_{e-1}$ and finally at the instant $e$ the tool $t$ was used to implement the job $T_e$. The elements of set $\mathscr{P}(L)$ we call pipes.

Along with the usual union symbol $\cup$ we use symbol $\sqcup$ for disjoint union, i.e. $\bigsqcup_{i}S_i$ means $\bigcup_{i}  S_i$, where for all $i \neq j$ $S_i \cap S_j = \varnothing$. 

\begin{lemma}\label{lemma-1}
Let $L \in \boldsymbol{L}$, then 
$$\displaystyle U(G_L) = \bigsqcup_{p \in \mathscr{H}_{0}(L)}  U(p) \ \sqcup  \ \bigsqcup_{p \in \mathscr{H}_{1}(L)} U(p) \  \sqcup \ \bigsqcup_{p \in \mathscr{P}(L)} U(p).$$
\label{lemma:1}
\end{lemma} 
\begin{proof}
Suppose that exists a pair of paths with common useless vertex, i.e. $ \exists p, p' \in \mathscr{H}_{0}(L) \cup \mathscr{H}_{1}(L) \cup \mathscr{P}(L):$ $p \neq p'$, $v_{k}^{t} \in U(p) \cap U(p') \neq \varnothing$. Arcs only connect vertices with the same tool, then $t'=t$ and $(v_{i}^{t})_{i=s}^{e} = p \neq p'=(v_{i}^{t})_{i=s'}^{e'}$.\\ 
Let $s<s'$, then $s' \leq k$. Since $p$ goes through $s,s+1, \dots , k$, then $t \in L_{s}, t \in L_{s+1}, \dots , t \in L_{s'-1}, \dots t \in L_{k}$. Note that $t \in L_{s'-1}$, then by the definitions of sets $\mathscr{H}_{0}(L)$ and $\mathscr{H}_{1}^{pre}(L)$, $p' \notin \mathscr{H}_{0}(L) \cup \mathscr{H}_{1}^{pre}(L)$. Then, $p' \in \mathscr{P}(L) \cup \mathscr{H}_{1}^{post}$ and then $t \in T_{s'}$. Vertex $v_{k}^{t}$ is useless, i.e. $v_{k}^{t} \in U(G_L)$ then $t \notin T_{k}$ (by the definition of $U(G_L)$) and then $s' \neq k$. Since, $s' \neq k$ and $s' \leq k$, then $s'<k$. Since $s'<k$ and $s<s'$ and $k \leq e$ then $s<s'<k \leq e$. Note that $t \notin \bigcup_{i=s+1}^{e-1} T_{i}$ by the definitions of sets $\mathscr{P}(L)$, $\mathscr{H}_{0}(L)$, $\mathscr{H}_{1}(L)$ and since $s<s'<e$, then $t \notin T_{s'}$, which contradicts with $t \in T_{s'}$. Thus $s \geq s'$. Similarly, can we obtain a contradiction with $s>s'$, then $s=s'$.

Let $e>e'$, then $e' \geq k$. Since $p$ goes through $k, k+1, \dots e$, then $t \in L_{k}, t \in L_{k+1}, \dots , t \in L_{e'+1}, \dots t \in L_{e}$. Note that $t \in L_{e'+1}$, then by the definitions of sets $\mathscr{H}_{0}(L)$ and $\mathscr{H}_{1}^{post}(L)$, $p' \notin \mathscr{H}_{0}(L) \cup \mathscr{H}_{1}^{post}(L)$. Since $p' \in \mathscr{P}(L) \cup \mathscr{H}_{1}^{pre}$, then $t \in T_{e'}$ and since $t \notin T_{k}$, then $e' \neq k$. Since, $e' \neq k$ and $e' \geq k$, then $e'>k$. Since $e'>k$ and $e>e'$ and $k \geq s$ then $e>e'>k \geq s$. Note that $t \notin \bigcup_{i=s+1}^{e-1} T_{i}$ by the definition of sets $\mathscr{P}(L)$, $\mathscr{H}_{0}(L)$, $\mathscr{H}_{1}(L)$ and since $e>e'>s$, then $t \notin T_{e'}$, which contradicts with $t \in T_{e'}$. Thus $e \leq e'$. Similarly, we obtain a contradiction with $e<e'$, then $e=e'$.\\
From $s=s'$ and $e=e'$ implies $p=p'$, which leads to a contradiction with assumption $p \neq p'$ and then $\forall p \neq p' \in \mathscr{H}_{0}(L) \cup \mathscr{H}_{1}(L) \cup \mathscr{P}(L)$ $U(p) \cap U(p') = \varnothing$. Consequentially $\bigcup_{p \in \mathscr{H}_{1}(L)}  U(p) \ \cup  \ \bigcup_{p \in \mathscr{H}_{0}(L)} U(p) \  \cup  \ \bigcup_{p \in \mathscr{P}(L)} U(p)  = \\ = \bigsqcup_{p \in \mathscr{H}_{1}(L)}  U(p) \ \sqcup  \ \bigsqcup_{p \in \mathscr{H}_{0}(L)} U(p) \  \sqcup  \ \bigsqcup_{p \in \mathscr{P}(L)} U(p)$.\\ 
Thus $\bigsqcup_{p \in \mathscr{H}_{1}(L)}  U(p) \ \sqcup  \ \bigsqcup_{p \in \mathscr{H}_{0}(L)} U(p) \  \sqcup  \ \bigsqcup_{p \in \mathscr{P}(L)} U(p) \subseteq U(G_L)$.

\begin{algorithm}
\SetAlgoLined
\DontPrintSemicolon
\SetKw{KwGoTo}{go to}
	$L_{0}:= \varnothing$, $L_{n+1} := \varnothing$\;
	$p:=(v_{k}^{t})$\;
	$s:=k$, $e:=k$\;
	\For{$i:=k-1, \dots , 1$}{
		\If{$t \in L_{i}$}{
			$p:= (v_{i}^{t}) \circ p$\;
			$s:=i$\;
		}
		\If{$t \in T_{i}$ or $t \notin L_{i-1}$}{
			\textbf{break}\;
		}
    }
    
	\For{$i:=k+1, \dots , n$}{
		\If{$t \in L_{i}$}{
			$p:= p \circ (v_{i}^{t}) $ \;
			$e:=i$\;
		}
		\If{$t \in T_{i}$ or $t \notin L_{i+1}$}{
			\textbf{break}\;
		}
    }
    
\
\textbf{return} $p,s,e$\;
  \caption{FindPath}
\end{algorithm}\label{algorithm:v_to_p}

In the first loop of the FindPath a vertex $v_{i}^{t}$ is added to the beginning of the path $p=(v_{j})_{j=i+1}^{e}$ only if $t \in L_i$. In the second loop of the FindPath a vertex $v_{i}^{t}$ is added to the end of the path $p=(v_{j})_{j=s}^{i-1}$  only if $t \in L_i$. Thus
\begin{equation}
t \in \bigcap_{i=s}^{e} L_{i}
\end{equation}

Both loops stop working if a point in time $i$ is found where the tool $t$ is required for the job, i.e. $t \in T_{i}$. Thus 
\begin{equation}
t \notin \bigcup_{j=s+1}^{e-1} T_{i}
\end{equation}
After FindPath execution, there are cases:
\begin{enumerate}
\item $t \in T_{s}$ and $t \in T_{e}$. Since (1) and (2) are satisfied, then all the conditions for $p \in \mathscr{P}(L)$ are satisfied.
\item $t \in T_{s}$ and $t \notin T_{e}$. From $t \notin T_{e}$ implies that, the second cycle either completed without breaks and then $e=n$, or the cycle was interrupted, when $t \notin L_{e+1}$. Since (1) and (2) are satisfied, then all the conditions for  $p \in \mathscr{H}_{1}^{post}(L)$ are satisfied.
\item $t \notin T_{s}$ and $t \in T_{e}$. From $t \notin T_{s}$ implies that, the first cycle either completed without breaks and then $s=1$, or the cycle was interrupted, when $t \notin L_{s-1}$. Since (1) and (2) are satisfied, then all the conditions for $p \in \mathscr{H}_{1}^{pre}(L)$  are satisfied.
\item $t \notin T_{s}$ and $t \notin T_{e}$. From $t \notin T_{s}$ implies that, the first cycle either completed without breaks and then $s=1$, or the cycle was interrupted, when $t \notin L_{s-1}$. 
From $t \notin T_{e}$ implies that, the second cycle either completed without breaks and then $e=n$, or the cycle was interrupted, when $t \notin L_{e+1}$. Since (1) and (2) are satisfied, then all the conditions for $p \in \mathscr{H}_{0}(L)$ are satisfied.
\end{enumerate}
 
Thus, for an arbitrary vertex $v_{k}^{t}$ such that $t \notin T_{k}$, i.e. $t \in U(G)$ found the path $p \in \mathscr{H}(L) \cup \mathscr{P}(L)$ which contains $v_{k}^{t}$. Then $U(G) \subseteq \bigsqcup_{p \in \mathscr{H}_{1}(L)}  U(p) \ \sqcup  \ \bigsqcup_{p \in \mathscr{H}_{0}(L)} U(p) \  \sqcup \  \bigsqcup_{p \in \mathscr{P}(L)} U(p)$. Then  $$U(G_L) = \bigsqcup_{p \in \mathscr{H}_{1}(L)}  U(p) \ \sqcup  \ \bigsqcup_{p \in \mathscr{H}_{0}(L)} U(p) \  \sqcup \  \bigsqcup_{p \in \mathscr{P}(L)} U(p) \ \Box. $$

\end{proof}

\begin{lemma}\label{lemma-2}
Let $L \in \boldsymbol{L}$, then
$$\displaystyle |A(G_L)| = \sum_{p \in \mathscr{P}(L)} |A(p)| + \sum_{p \in \mathscr{H}_{0}(L)} |A(p)| + \sum_{p \in \mathscr{H}_{1}(L)} |A(p)|. $$
\label{lemma:2}
\end{lemma} 

\begin{proof}
Since $\displaystyle V(G_L) \supseteq V(\mathscr{H}_{0}(L)) \cup V(\mathscr{H}_{1}(L)) \cup V(\mathscr{P}(L))$, then\\ $\displaystyle A(G_L) \supseteq A(\mathscr{H}_{0}(L)) \cup A(\mathscr{H}_{1}(L)) \cup A(\mathscr{P}(L))$.\\
Let $a = (v_{i}^{t},v_{i+1}^{t}) \in A(G_L)$, then there are cases:
\begin{itemize}

\item $t \in T_i$ and $t \in T_{i+1}$. Then $t \in T_i \cap T_{i+1}$, $t \notin \bigcup_{i+1}^{i+1-1} T_{i}=\varnothing$, $t \in L_i \cap L_{i+1}$, then $p_{i,i+1}^{t} = (v_{i}^{t},v_{i+1}^{t}) \in \mathscr{P}(L)$.
\item $t \notin T_{i}$ or $t \notin T_{i+1}$. Let $j \in \{i,i+1\}: t \notin T_{j}$, then according to Lemma~\ref{lemma:1} $v_{j}^{t} \in U(G_L) = \bigsqcup_{p \in \mathscr{H}_{0}(L)}  U(p) \ \sqcup  \ \bigsqcup_{p \in \mathscr{H}_{1}(L)} U(p) \  \sqcup \ \bigsqcup_{p \in \mathscr{P}(L)} U(p)$, then arc $a$ is part if only one path from $\mathscr{H}_{0}(L) \sqcup \mathscr{H}_{1}(L) \sqcup \mathscr{P}(L)$.

\end{itemize}
Then $A(G_L) = \bigsqcup_{p \in \mathscr{H}_{0}(L)}  A(p) \ \sqcup  \ \bigsqcup_{p \in \mathscr{H}_{1}(L)} A(p) \  \sqcup \ \bigsqcup_{p \in \mathscr{P}(L)} A(p)$, then $$|A(G_L)| = \sum_{p \in \mathscr{P}(L)} |A(p)| + \sum_{p \in \mathscr{H}_{0}(L)} |A(p)| + \sum_{p \in \mathscr{H}_{1}(L)} |A(p)| \ \Box .$$
\end{proof}
\begin{lemma}\label{lemma-3}
$L \in \boldsymbol{L}$, $M = ToFullMag(L)$, then
\begin{enumerate}
\item $M \in \boldsymbol{M}$
\item $\mathscr{H}_{0}(L) = \varnothing \implies \mathscr{H}_{0}(M) = \varnothing$
\item $|\mathscr{P}(L)| \leq |\mathscr{P}(M)|$
\end{enumerate}
\label{lemma:3}
\end{lemma} 

\begin{proof}
Let $L'$ be the result of the first half of ToFullMag iterations, when  $$(i,j) = (1,2), (2,3), \dots, (n-1,n).$$
Let us prove that $|L_{n}'|=C$. 
Assume the opposite $|L_{n}'|<C$. 
Note that ToFullMag adds tools form $L_{n-1}'$ to $L_n$ while $L_n'=C$ and $L_{n-1}' \subseteq L_{n}'$ are false. By the assumption $L_n' \neq C$, then $L_{n-1}' \subseteq L_{n}'$. From $L_{n-1}' \subseteq L_{n}'$ and $|L_{n}'|<C$ implies $|L_{n-1}'|<C$. 
Consider the previous iteration of the algorithm, when tools from $L_{n-2}'$  are added to $L_{n-1}$. 
Then, similarly to the previous reasoning $L_{n-2}' \subseteq L_{n-1}'$. 
Continuing the reasoning, we get $L_{1}' \subseteq L_{2}' \subseteq \dots \subseteq L_{n}'$, consequently $\bigcup_{i=1}^{n} L_{n}' = L_{n}'$. 
Since ToFullMag only adds tools in states, then for all $i$ $|\bigcup_{i=1}^{n} L_{n}| \leq |\bigcup_{i=1}^{n} L_{i}'|$, then $|\bigcup_{i=1}^{n} L_{i}| \leq |\bigcup_{i=1}^{n} L_i'|$. 
By the definition of $\boldsymbol{L}$, $T_1 \subseteq L_1 ,\dots T_n \subseteq L_n$, then$|\bigcup_{i=1}^{n} T_{i}| \leq |\bigcup_{i=1}^{n} L_i|$. Then $m = |\bigcup_{i=1}^{n} T_{i}| \leq |\bigcup_{i=1}^{n} L_i| \leq |\bigcup_{i=1}^{n} L_i'| = |L_n'|<C$, then $m<C$ but by the $TLP$ condition $m > C$ is satisfied, we got a contradiction, then $|L_{n}'|=C$.\\
Let $L''$ be the result of the second half of the ToFullMag iterations, then $(i,j) = (n,n-1), (n-1,n-2), \dots, (2,1)$.
Since $L \in \boldsymbol{L}$, then $|L_1| \leq C, \dots |L_n| \leq C$. Since ToFullMag adds tools to the state at time $j$ only if the magazine is not yet full at time $j$, then $|L_1''| \leq C, \dots |L_n''| \leq C$.
In the second half of the ToFullMag iterations adds elements from the next state $L_{j+1}''$ to previous state $L_{j}'$, while  $|L_{j}''|=C$ and $L_{j+1}'' \subseteq L_{j}''$ are false, then either $|L_{n-1}''|=C$ or $L_{n}'' \subseteq L_{n-1}''$ is true, then $|L_{n-1}''|=C$. 
Similarly, for next iterations, we obtain  that $|L_{1}''|=|L_{2}''|= \dots =|L_{n}''|=C$, and consequently $L'' \in \boldsymbol{M}$, thus $L'' = ToFullMag(L)  =: M \in \boldsymbol{M}$.\\
Let us prove that $\mathscr{H}_{0}(L) = \varnothing \implies \mathscr{H}_{0}(M) = \varnothing$. 
Assume the opposite $p = (v_{i}^{t})_{i=s}^{e} \in \mathscr{H}_{0}(M)$. $\mathscr{H}_{0}(L)= \varnothing$, then while ToFullMag execution, at least one of the vertices $V(p)$ has been added to $G_L$. Let $v_{i}^{t}$ be the first of the added vertices. 
Note that $v_{i}^{t} \in V(p)$, $p \in \mathscr{H}_{0}(M)$, then $t \notin T_{i-1},t \notin T_{i},t \notin T_{i+1}$ according to the definition of $\mathscr{H}_{0}$. 
Then if $v_{i}^{t}$ was added in the first half of iterations ToFullMag, then $t \in L'_{i-1}$, where either $v_{i-1}^{t} \in T_{i-1}$, which contradicts with $v_{i-1}^{t} \notin T_{i-1}$, or $v_{i-1}^{t} \in U(p)$ and consequently $v_{i-1}^{t} \in U(p)$. Since $v_{i}^{t}$ is the first of the added vertices, then $v_{i-1}^{t}$ was already in $G_L$. $v_{i-1}^{t}$ was in $G_{L}$ and $\mathscr{H}_{0}(L) = \varnothing$, then $v_{i-1}^{t} \notin \bigcup_{p \in \mathscr{H}_{0}(L)} U(p)$, then according to Lemma~\ref{lemma:1} we have $v_{i-1}^{t} \in \bigsqcup_{p \in \mathscr{H}_{1}(L)} U(p) \  \sqcup \ \bigsqcup_{p \in \mathscr{P}(L)} U(p)$. 
Vertex $v_{i-1}^{t}$ was in $V(G_L)$ and $v_{i}^{t}$  was not, then exists some path in $G_L$ that ends with $v_{i-1}^{t}$, i.e. $\exists p'=(v_{s'}^{t}, \dots , v_{i-1}^{t}) \in \mathscr{H}_{0}(L) \cup\mathscr{H}_{1}(L) \cup \mathscr{P}(L) = \mathscr{H}_{1}(L) \cup \mathscr{P}(L)$. 
Since $t \notin L_{i-1}$, then  $p' \notin \mathscr{H}_{1}^{pre}(L) \cup \mathscr{P}(L)$ by definitions of sets $\mathscr{H}_{1}^{pre}, \mathscr{P}$.
Then $p' \in \mathscr{H}_{1}^{post}$, but then $p=p' \circ (v_{i}^{t}, v_{i+1}^{t}, \dots v_{e}^{t}) =(v_{s'}^{t},\dots v_{i+1}^{t}, \dots v_{e}^{t})$. 
Since $p' \in \mathscr{H}_{1}^{post}$, then $t \in L_{s'}$ and then $t \in M_{s'}$, which contradicts with $p \in \mathscr{H}_{0}(M)$ by the definition of $\mathscr{H}_{0}$.
We have obtained a contradiction with the assumption. 
Similarly, (from the symmetry of ToFullMag) for the case when $v_{i}^{t}$ was constructed in the second half of the ToFullMag iterations, we get a contradiction. 
Thus $\mathscr{H}_{0}(L) = \varnothing \implies \mathscr{H}_{0}(L) = \varnothing$\\
Since ToFullMag only adds vertices on $G_L$, then  $\mathscr{P}(L) \subseteq \mathscr{P}(M)$ and finally $|\mathscr{P}(L)| \leq |\mathscr{P}(M)|$ $\Box$.
\end{proof}
Let us prove Theorem 1.
\begin{proof}
Let $p=(v_{s}^{t}, \dots, v_{e}^{t} ) \in \mathscr{P}(M)$, then exactly two vertices do not belong to $U(p)$, these are $v_{s}^{t}$, $v_{e}^{t}$, then $|V(p)|=|U(p)| + 2$.\\
According to Lemma~\ref{lemma:1} we have $U(\mathscr{P}(M)) = \bigsqcup_{\mathscr{P}(M)} U(p)$, then $|U(\mathscr{P}(M))| = \sum_{\mathscr{P}(M)} |U(p)|$.\\
Then $\sum_{p \in \mathscr{P}(M)} |A(p)| = \sum_{p \in \mathscr{P}(M)} (|V(p)|-1) = \sum_{p \in \mathscr{P}(M)} (|U(p)|+2-1) = \\ = \sum_{p \in \mathscr{P}(M)} (|U(p)|+1) = \sum_{p \in \mathscr{P}(M)} |U(p)| + |\mathscr{P}(M)| = |U(\mathscr{P}(M))| +  |\mathscr{P}(M)|$.
\begin{equation}
\sum_{p \in \mathscr{P}(M)} |A(p)| = |U(\mathscr{P}(M))| + |\mathscr{P}(M)|
\end{equation}

Let $p=(v_{s}^{t}, \dots, v_{e}^{t} ) \in \mathscr{H}_{0}(M)$.Then, $V(p) = U(p)$ and then $|V(p)|=|U(p)|$.\\
According to Lemma~\ref{lemma:1} we have $U(\mathscr{H_{0}}(M)) = \bigsqcup_{\mathscr{H}_{0}(M)} U(p)$, then $|U(\mathscr{H}_{0}(M))| = \sum_{\mathscr{H}_{0}(M)} |U(p)|$.\\
Then $\sum_{p \in \mathscr{H}_{0}(M)} |A(p)| = \sum_{p \in \mathscr{H}_{0}(M)} (|V(p)|-1) = \sum_{p \in \mathscr{H}_{0}(M)} (|U(p)|-1) = \\ = \sum_{p \in \mathscr{H}_{0}(M)} |U(p)| -|\mathscr{H}_{0}(M)| = |U(\mathscr{H}_{0}(M))| -|\mathscr{H}_{0}(M)|$.
\begin{equation}
\sum_{p \in \mathscr{H}_{0}(M)} |A(p)| = |U(\mathscr{H}_{0}(M))| -|\mathscr{H}_{0}(M)|
\end{equation}

Let $p=(v_{s}^{t}, \dots, v_{e}^{t} ) \in \mathscr{H}_{1}(M)$, then exactly one vertex does not belong to $U(p)$, this is either $v_{s}^{t}$ or $v_{e}^{t}$, then $|V(p)|=|U(p)| + 1$.
According to Lemma~\ref{lemma-1} we have $U(\mathscr{H_{1}}(M)) = \bigsqcup_{\mathscr{H}_{1}(M)} U(p)$, then $|U(\mathscr{H}_{1}(M))| = \sum_{\mathscr{H}_{1}(M)} |U(p)|$.
Then $\sum_{p \in \mathscr{H}_{1}(M)} |A(p)| = \sum_{p \in \mathscr{H}_{1}(M)} (|V(p)|-1) = \sum_{p \in \mathscr{H}_{1}(M)} (|U(p)|+1-1) = \sum_{p \in \mathscr{H}_{1}(M)} |U(p)| = |U(\mathscr{H}_{1}(M))|$.
\begin{equation}
\sum_{p \in \mathscr{H}_{1}(M)} |A(p)| = |U(\mathscr{H}_{1}(M))|
\end{equation}
According to Lemma~\ref{lemma:2} we have 
$\displaystyle |A(G_M)| = \sum_{p \in \mathscr{P}(M)} |A(p)| + \sum_{p \in \mathscr{H}_{0}(M)} |A(p)| + \sum_{p \in \mathscr{H}_{1}(M)} |A(p)| = |U(\mathscr{P}(M))| + |\mathscr{P}(M)| + |U(\mathscr{H}_{0}(M))| -|\mathscr{H}_{0}(M)| + |U(\mathscr{H}_{1}(M))| = |U(G_M)| + |\mathscr{P}(M)|  -|\mathscr{H}_{0}(M)|$.

$switeches(M) = (n-1)C - |A(G_M)| = C(n-1) - (|U(G_M)| + |\mathscr{P}(M)|  -|\mathscr{H}_{0}(M)|) = Cn - C - (Cn - \sum_{i=1}^{n} |T_i| + |\mathscr{P}(M)|  -|\mathscr{H}_{0}(M)|) = \sum_{i=1}^{n} |T_i| - C - |\mathscr{P}(M)|  +|\mathscr{H}_{0}(M)|$.\\
Note that $\sum_{i=1}^{n} |T_i|$ and $C$ are given by TLP, and they can not change, then $switeches(M)$ depends on $|\mathscr{P}(M)|$ and $|\mathscr{H}_{0}(M)|$ only. Note that if for some $M$ the number of pipes $|\mathscr{P}(M)|$ is maximized and for the same $M$ set $\mathscr{H}_{0}(M)$ is empty, then $switches(M)$ is minimized.\\
Let $M \in \boldsymbol{M}$ such that $|\mathscr{P}(M)| = max\{|\mathscr{P}(M'')|: M'' \in \boldsymbol{M}\}$, 
$\mathscr{H}_{0}(M) \neq \varnothing$.  Since $V(\mathscr{H}_{0}(M)) = U(\mathscr{H}_{0}(M))$ 
and according to Lemma~\ref{lemma:1} $U(\mathscr{H}_{0}(M)) \cap U(\mathscr{P}(M))=\varnothing$, 
then we can delete $V(\mathscr{H}_{0}(M))$ from $G_M$ with no decreasing of $|\mathscr{P}(M)|$ and we get $L \in \boldsymbol{L}$ as a result of deletion, where $|\mathscr{P}(L)| = |\mathscr{P}(M)|$. Let $M' = ToFullMag(L)$, according to Lemma~\ref{lemma:3} $M' \in \boldsymbol{M}, \mathscr{H}_{0}(M') = \varnothing, |\mathscr{P}(L)| \leq |\mathscr{P}(M')|$, then $max\{|\mathscr{P}(M'')|: M'' \in \boldsymbol{M}\} = |\mathscr{P}(M)| \leq |\mathscr{P}(M')|$, consequently $max\{|\mathscr{P}(M'')|: M'' \in \boldsymbol{M}\} = |\mathscr{P}(M')|$.
Finally $switches(M') = \sum_{i=1}^{n}|T_i| -C - \underset{M'' \in \boldsymbol{M}}{\mathrm{max}}\{|\mathscr{P}(M'')|\}$ and then $\underset{M'' \in \boldsymbol{M}}{\mathrm{min}}\{switches(M'')\} = \sum_{i=1}^{n}|T_i| -C - \underset{M'' \in \boldsymbol{M}}{\mathrm{max}}\{|\mathscr{P}(M'')|\}  $ $\Box .$
\end{proof}
\subsection{Proof of Theorem 2} \label{section:Proof of Theorem 2}
Let $\Pi(\boldsymbol{L})= \{\pi_{s,e}^{t}: \exists L \in \boldsymbol{L}:\pi_{s,e}^{t} \in \mathscr{P}(L)\}$ be the set of all possible pipes. \\
An adding of tool $t$ to the magazine states $L_{s+1}, \dots , L_{e-1}$ we call construction of the pipe $\pi_{s,e}^{t}$ in $L$ for given $L \in \boldsymbol{L}$ and $\pi_{s,e}^{t} \in \Pi(\boldsymbol{L}) \backslash \mathscr{P}(L)$.\\
Let $\Pi(L) = \{\pi_{s,e}^{t} \in \Pi(\boldsymbol{L}) \backslash \mathscr{P}(L): \forall j \in \{s+1,\dots ,e-1\} \ t \notin L_j,|L_j|<C \}$ be the set of all possible pipes $\pi_{s,e}^{t}$ that can be constructed (by adding the tool $t$ to the magazine states $L_{s+1}, \dots , L_{e-1}$).

Let $ \boldsymbol {\widehat {L}} = \{L \in \boldsymbol{L}: U(L)=U(\mathscr{P}(L))\}$ be the set of all possible sequences of states, in which empty slots were filled only by pipes or remained empty. In other words, all useless vertexes belong to pipes.\\
The following lemma shows that a pipe $\pi_{s,e}^{t}$ can be constructed iff it has not been constructed yet and there are enough empty slots at instants $s+1, \dots e-1$.
\begin{lemma}\label{lemma-4}
Let $L \in \boldsymbol{\widehat{L}}$, then $$\Pi(L) = \{\pi_{s,e}^{t} \in \Pi(\boldsymbol{L})\backslash \mathscr{P}(L): \forall j \in \{s+1, \dots , e-1\}  \ |L_j|<C \}.$$
\label{lemma:4}
\end{lemma}
\begin{proof}

Based on the definition of $\Pi(L)$, it suffices to prove that if $L \in \boldsymbol{\widehat{L}}$ and $\pi_{s,e}^{t} \in \Pi(\boldsymbol{L})\backslash \mathscr{P}(L)$, then $\forall j \in \{s+1, \dots , e-1\}  $ $ t \notin L_j$. 
Suppose $\exists j \in \{s+1, \dots , e-1 \}: t \in L_j$, since $\pi_{s,e}^{t}$ is a pipe, then $\forall i \in \{s+1,\dots , e-1 \}  $ $ t \notin T_i$, then $t \notin T_j$, which implies that vertex $v_{j}^{t} \in U(L)$ by the definition of $U$. 
Since $L \in \boldsymbol{\widehat{L}}$ then $U(L) = U(\mathscr{P}(L))$ and then $\exists \pi_{s',e'}^{t} \in \mathscr{P}(L): s' < j < e'$. Since $\pi_{s',e'}^{t}$ is a pipe, then $\forall i \in \{s'+1,\dots , e'-1 \}  $ $ t \notin T_i$, and since $\pi_{s,e}^{t}$ is a pipe, then $ $ $ t \in T_s$, then $s \leq s'$. 
Since $\pi_{s,e}^{t}$ is a pipe, then $\forall i \in \{s+1,\dots , e-1 \}  $ $ t \notin T_i$, and since $\pi_{s',e'}^{t}$ is a pipe, than $ $ $ t \in T_{s'}$ than $s' \leq s$, then $s=s'$ and similarly $e=e'$, which implies that $\pi_{s',e'}^{t}$ and $\pi_{s,e}^{t}$ are the same pipe, then $\pi_{s,e}^{t} \in \mathscr{P}(L)$ and $\pi_{s,e}^{t} \in \Pi(\boldsymbol{L}) \backslash \mathscr{P}(L)$. Then, $\pi_{s,e}^{t} \in (\Pi(\boldsymbol{L}) \backslash \mathscr{P}(L)) \cap \mathscr{P}(L) = \varnothing$ which leads to a contradiction $\Box$.\\
\end{proof}
\noindent Further, we always assume that $ L \in \boldsymbol {\widehat {L}}  $, since we will only talk about constructing and removing pipes. \\
\noindent  Let $ \boldsymbol{L}_{opt} = \{L \in \boldsymbol {\widehat {L}}: | \mathscr{P} (L) | = max \{| \mathscr{P} (M) |: M \in \boldsymbol {M} \} \} $, i.e. this is the set $ L \in \boldsymbol {\widehat {L}}  $: $ L $ contains the largest possible number of pipes. \\
Let $L \in \boldsymbol {\widehat {L}}, K \subseteq \Pi(\boldsymbol{L})$. By the deletion of $K$, we mean deletion of vertexes $U(K)$ from $G_L$, i.e. we empty all slots which are not used for implementing jobs that occupied by pipes from $K$. Note that if $K=\mathscr{P}(L)$, then $U(L)$ will be deleted, and then we get $L^{min}:=(T_1,T_2, \dots , T_n)$ as a result of deletion. Similarly, by the constructing of $K'$, we mean adding of vertexes $U(K')$ to $G_L$.
\noindent $\mathscr{K}(L):=\{(K,K') \subseteq \Pi(\boldsymbol{L})^2: |K|<|K'|$, it is possible to remove $K$ then construct $K'$ in $L\}$, where $L \in \widehat{L}$. 
\noindent $\mathscr{K}_{min}(L): = \{(K, K ') \in \mathscr{K}(L): \nexists (\widetilde{K}, \widetilde{K}' ) \in \mathscr{K} (L): (\widetilde{K} \subseteq K $, $ \widetilde{K}' \subset K ')$ or $(\widetilde{K} \subset K $, $ \widetilde{K}' \subseteq K ') \} $

\begin{lemma}\label{lemma-5}
Let $L \in \boldsymbol{\widehat{L}}$, then $\mathscr{K}_{min}(L)=\varnothing \iff L \in \boldsymbol{L}_{opt}$.
\label{lemma:5}
\end{lemma}
\begin{proof}
Let us first prove $\mathscr{K}(L)=\varnothing \iff L \in \boldsymbol{L}_{opt}$.\\
If $L$ is not optimal and the empty slots were not filled with anything other than constructing pipes, then we can remove all pipes from it, freeing the occupied slots that are not needed to implement the jobs and construct more pipes than were removed. 
Let $L \in \boldsymbol{\widehat{L}}, L \notin \boldsymbol{L}_{opt}$, then $\exists L' \in \boldsymbol{\widehat{L}}: |\mathscr{P}(L)| < |\mathscr{P}(L')|$. 
Let $K:=\mathscr{P}(L)$, $K':=\mathscr{P}(L')$. After removing $K$ from $L$, we get $L^{min}$ as a result. 
Since $L' \in \boldsymbol{\widehat{L}}$, then $L^{min}_{1} = T_1 \subseteq L_1', L^{min}_{2} = T_2 \subseteq L_2', \dots, L^{min}_{n} = T_n \subseteq L_n'$. 
Then it is possible to construct $K'$ in $L^{min}$, thereby constructing all pipes from $K'$ and get $L'$ as a result. 
Thus, $\exists (K,K'): |K|<|K'|$ it is possible to remove $K$ and construct $K'$, then $(K,K') \in \mathscr{K}(L)$ and then $\mathscr{K}(L) \neq \varnothing$.\\
Let $(K,K') \in \mathscr{K}(L) \neq \varnothing$. Let  $L'' \in \boldsymbol{\widehat{L}}$ be the result of deletion of $K$ and constructing $K'$ in $L$. Since $|K|<|K'|$, then $|\mathscr{P}(L)|<|\mathscr{P}(L'')|$, consequently $ L \notin \boldsymbol{L}_{opt}$.\\
From the above, it follows that $\mathscr{K}(L) \neq \varnothing \iff L \notin \boldsymbol{L}_{opt}$. Then $\mathscr{K}(L) = \varnothing \iff L \in \boldsymbol{L}_{opt}$.\\
Note that $\mathscr{K}_{min}(L) \subseteq \mathscr{K}(L)$, then $\mathscr{K}(L)=\varnothing \implies \mathscr{K}_{min}(L) =\varnothing$.\\
Let $\mathscr{K}_{min}(L) = \varnothing$ and $(K,K') \in \mathscr{K}(L)$. If $\nexists (\widetilde{K}, \widetilde{K}' ) \in \mathscr{K} (L): (\widetilde{K} \subseteq K $, $ \widetilde{K}' \subset K ')$ or $(\widetilde{K} \subset K $, $ \widetilde{K}' \subseteq K ')$, then $(K,K') \in \mathscr{K}_{min}(L)$, which leads to a contradiction, but if exists $(\widetilde{K}, \widetilde{K}' )$ then let us rename $K:= \widetilde{K}$, $K' = \widetilde{K}'$ and repeat the same reasoning. $K,K'$ are finite sets and with each renaming either $K$ or $K'$ decrease their cardinality.  
Then at some renaming there will be no pair $(\widetilde{K},\widetilde{K}')$ or $K=K'=\varnothing$. Note that for $K=K'=\varnothing$ there is no pair $(\widetilde{K},\widetilde{K}')$ either, then $(K,K') \in \mathscr{K}_{min}(L)$, which leads to a contradiction. Consequently $\mathscr{K}_{min}(L) = \varnothing \implies \mathscr{K}(L) = \varnothing$. Then $\mathscr{K}_{min}(L) = \varnothing \iff \mathscr{K}(L) = \varnothing$. Consequently $\mathscr{K}(L)=\varnothing \iff \mathscr{K}_{min}(L)=\varnothing \iff L \in \boldsymbol{L}_{opt}$ $\Box$.
\end{proof}
Let $ N(\pi_{s,e}^{t}) = \{s + 1, s + 2, \dots, e-1 \} $, i.e. the set of all times at which one empty slot is needed to construct the pipe $ \pi_{s,e}^{t} $.
Let $\displaystyle N(K) = \bigcup _ {\pi_{s,e}^{t} \in K} N(\pi_{s,e}^{t})$. 

\begin{lemma}\label{lemma-6}
Let $L \in \boldsymbol{\widehat{L}}, (K, K ') \in \mathscr{K} (L), \pi \in K, \tau \in K'$, then $$N(\pi) \cap  N(K') \subseteq N(\tau)  \implies (K,K') \notin \mathscr{K}_{min}(L).$$
\label{lemma:6}
\end{lemma}

\begin{proof}
Let $ (K, K ') \in \mathscr{K} (L), \pi \in K, \tau \in K '$, $N(\pi) $ $\cap$ $ N(K') \subseteq N(\tau)$. \\
Let $\pi = (v_{s}^{t},v_{s+1}^{t}, \dots ,v_{e}^{t})  \in K'$, then $(K \backslash \{\pi\}, K' \backslash \{\pi\}) \in \mathscr{K}(L)$. $K \backslash \{\pi\} \subset K, K' \backslash \{\pi\} \subset K'$, then by definition of  $\mathscr{K}_{min}$ condition $(K,K') \notin \mathscr{K}_{min}(L)$ will be satisfied. Let $\pi \notin K'$. Let $L'$ be the result of deletion $K$ and constructing $K'$ in $L$. Note that after removing $\pi$ at every moment in $N(\pi)$ will be at least one empty slot. Then after removing remaining $N(K) \backslash N(\pi)$ at every moment in $N(\pi)$ will be at least one empty slot too. And after construction of all pipes from $K'$ at all times in $N(\pi) \backslash N(K')$ exists at least one empty slot in $L'$. Let $L''$ be the result of deletion $\tau$ from $L'$. At all the instants from $N(\tau)$ exists at least one empty slot in $L''$. Since $N(\pi) $ $\cap$ $ N(K') \subseteq N(\tau)$, then at all times in $N(\pi) $ $\cap$ $ N(K')$ exists at least one empty slot too. At all times in $N(\pi) \backslash N(K')$ and $N(\pi) \cap N(K')$ exists at least one empty slot, then at all times in $N(\pi)$ exists at least one empty slot, i.e. $\forall j \in \{s+1,\dots,e-1\}$ $|L_j''|<C$. Since $\pi \in K, \pi \notin K'$, then $\pi \notin \mathscr{P}(L'') = ((\mathscr{P}(L) \backslash K) \cup K') \backslash \{\tau\}$. Since $\pi \in \Pi(\boldsymbol{L})\backslash \mathscr{P}(L'')$ and $\forall j \in \{s+1, \dots , e-1\}  $ $ |L_j''|<C \}$, according to Lemma~\ref{lemma:4} $\pi \in \Pi(L'')$. Let $L'''$ be the result of constructing $\pi$ in $L''$. Let $\widetilde {K} = K \backslash \{\pi\}$, $\widetilde {K}' = K' \backslash \{\tau\}$. Note that $L'''$ is result of deletion $\widetilde {K}$ from $L$ and constructing $\widetilde{K}'$. $|K| = \widetilde {K} + 1$, $|K'| = \widetilde {K}' + 1$ and since $|K|<|K'|$ then  $|\widetilde {K}| < |\widetilde {K'}|$. By definition of $\mathscr{K}$ condition $(\widetilde {K},\widetilde {K}') \in \mathscr{K}(L)$ will be satisfied. Note that $\widetilde {K} \subset K$, $\widetilde{K}' \subset K'$, then since $(\widetilde {K},\widetilde {K}') \in \mathscr{K}(L)$, then by definition of $\mathscr{K}_{min}$ condition $(K,K') \notin \mathscr{K}_{min}(L)$ will be satisfied $\Box$.
\end{proof}

Let us prove Theorem 2.

\begin{proof}
According to Lemma~\ref{lemma:3} and Lemma~\ref{lemma:5}, it will suffice to prove that
$\mathscr{K}_{min}(L)=\varnothing$, where $L=GPCA(T_1,\dots ,T_n;C)$.\\
Suppose the opposite, i.e. $\exists (K,K') \in \mathscr{K}_{min}(L)$. Since the Algorithm~\ref{algorithm:1} tries to construct all pipes from $ \Pi(\boldsymbol{L}) $ and constructs if possible. Then it is impossible to construct one more pipe in $ L $ without deleting one before this, therefore the set of pipes to be removed is always not empty, so $ K \neq \varnothing $.  $ | K '|> | K | $ by definition, then $ K' \neq \varnothing $.\\
\textbf{Case 1}: $e = min\{ e : \pi_{s,e}^{t} \in K \} \leq min\{ e : \pi_{s,e}^{t} \in K' \}$.\\
Then let $ \pi_{s,e}^{t} \in K $, where $s$, $t$ chosen arbitrarily; $s'=min\{i: \pi_{i,j}^{k} \in K' \}$, $ \pi_{s',e'}^{t'} \in K' $, where $e'$, $t'$ chosen arbitrarily. Since $ s'$ is minimal and $ e  \leq e'$ it follows that $N(\pi_{s,e}^{t}) 
$ $\cap$ $ N(K') \subseteq N(\pi_{s',e'}^{t'})$. Therefore, according to Lemma~\ref{lemma:6} $(K,K') \notin \mathscr{K}_{min}(L)$, which leads to a contradiction.

\textbf{Case 2}: $min\{ e : \pi_{s,e}^{t} \in K \} > min\{ e : \pi_{s,e}^{t} \in K' \}=e'$.\\
Then let $\pi_{s',e'}^{t'} \in K'$, where $s'$, $t'$ chosen arbitrarily; $s=min\{i: \pi_{i,j}^{k} \in K, j=e \}$, $\pi_{s,e}^{t} \in K$, where $e$, $t$ chosen arbitrarily.

Emptying the slots at each of the times from the set $ N(\pi_{s,e}^{t}) $ will allow us to construct a pipe $ \pi_{s',e'}^{t'} $, otherwise even after removing $ K $ it would not be possible to construct the pipe $ \pi_{s',e'}^{t'} $. Then after removing the pipe $ \pi_{s,e}^{t} $ it will be possible to construct $ \pi_{s',e'}^{t'} $. Consequently, the Algorithm 1 constructed the pipe $ \pi_{s',e'}^{t'} $ at an iteration earlier than $ \pi_{s,e}^{t} $, since $ e' <e $, and the Algorithm 1 iterates over the variable $ e $ in ascending order. At the iteration, when the variable $ e $ was equal to $ e '$, the pipe $ \pi_{s',e'}^{t'} $ should have been built, since at that moment the pipes $ \pi_{s,e}^{t} $ didn't exist yet, which is the same as it was removed. From which it follows that $ \pi_{s',e'}^{t'} $ has already been built. Let's move on to  $ (\widetilde {K}, \widetilde {K} '): \widetilde {K} = K \backslash \{\pi_{s',e'}^{t'} \} $, $ \widetilde {K} '= K' \backslash \{\pi_{s',e'}^{t'} \} $ and note that after removing $ \widetilde {K} $ it will be possible to construct all pipes from $ \widetilde {K } '$, i.e. $ (\widetilde {K}, \widetilde {K} ') \in \mathscr{K} (L) $, but $ \widetilde {K} \subset K $, $ \widetilde {K}' \subset K ' $. Then, by the definition of $ \mathscr{K}_{min}$, $ (K, K ') \notin \mathscr{K}_{min}(L) $, which leads to a contradiction.

\textbf {Case 1}, \textbf {Case 2} led to a contradiction, hence the assumption about $ \exists (K, K ') \in \mathscr{K}_{min} (GPCA(T_1 , \dots , T_n;C)) $ is not true, then by Lemma~\ref{lemma:5} we have $$ GPCA(T_1 , \dots , T_n;C) \in \boldsymbol {L_ {opt}},$$ which implies $$|\mathscr{P}(GPCA(T_1,\dots ,T_n;C))| = max\{ |\mathscr{P}(M)|: M \in \boldsymbol{M} \}.$$ Applying Theorem 1 we have 
$$ min\{switches(M''):M'' \in \boldsymbol{M}\}  = \sum_{i=1}^{n}|T_i| -C - |\mathscr{P}(GPCA(T_1,\dots ,T_n;C))| , $$ which proves the first statement of Theorem 2. 

Let $M=ToFullMag(GPCA(T_1,\dots ,T_n;C))$. Applying Lemma~\ref{lemma:3} we have $|\mathscr{P}(M)| \geq |\mathscr{P}(GPCA(T_1,\dots ,T_n;C))|$ and $M \in \boldsymbol{M}$, then according to the first statement of Theorem 2 we have  $$ToFullMag(GPCA(T_1,\dots ,T_n;C)) \in argmin\{switches(M): M \in \boldsymbol{M}\}
 \ \Box . $$
\end{proof}

\end{document}